\documentclass[prd,showpacs,preprintnumbers,amsmath,amssymb]{revtex4}

\usepackage{graphicx}
\usepackage{dcolumn}
\usepackage{bm}

\begin{document}

\title{QED Effective Actions in Space-Dependent Gauge and Electromagnetic Duality}

\author{Sang Pyo Kim}
\affiliation{Department of Physics, Kunsan National University, Kunsan 573-701, Korea}
\affiliation{Institute of Astrophysics, Center for Theoretical Physics,Department of Physics, National Taiwan University, Taipei 106, Taiwan}
\affiliation{Yukawa Institute for Theoretical Physics, Kyoto University, Kyoto 606-8502, Japan} \email{sangkim@kunsan.ac.kr}

\medskip

\date{\today}

\begin{abstract}
We develop the in-out formalism for one-loop effective actions in electromagnetic fields in the space-dependent gauge. We further advance a method using the inverse scattering matrix to calculate the effective actions in pure magnetic fields, find the effective actions in a constant magnetic field and a localized Sauter-type magnetic field and apply the uniform semiclassical approximation to the effective action in a general magnetic field. In the in-out formalism we show that the one-loop effective actions in constant fields and Sauter-type fields exhibit the electromagnetic duality.
\end{abstract}

\maketitle

\section{Introduction}\label{introduction}

The effective action in a background field can probe the vacuum structure of the underlying theory.
In quantum electrodynamics (QED) the effective action in a constant electromagnetic field was first found by Heisenberg, Euler, and
Schwinger in spinor QED \cite{heisenbergeuler} and by Weisskopf and Schwinger in scalar QED \cite{weisskopf,schwinger}.
The vacuum polarization by an electromagnetic field leads to prominent phenomena such as photon splitting \cite{adler},
direct photon-photon scattering \cite{erber}, birefringence \cite{klni}, and Schwinger pair production.
QED effective action since then has been an issue of constant interest and continuous investigations since it provides a field theoretical framework
for testing nonperturbative methods and for understanding the vacuum structure. Nonperturbative aspects of QED effective action
may be found, for instance, in Refs. \cite{dittrichreuter,fgs,dittrichgies,schubert-rev,dunne-rev}.

Recently, QED effective action has brought an extensive study since strong laser sources such as Extreme Light Infrastructure (ELI)
may observe direct photon-photon scattering and probably electron-positron pair production in the near future \cite{ringwald,mash,dunne-work},
not to mention astrophysical implications \cite{behpt,ruffini}. However, computing nonperturbative effective action beyond a constant
electromagnetic field has been considered a nontrivial, challenging task and effective actions have been known only for few configurations
of electromagnetic fields \cite{dunne-rev}. For this and other reasons most literature on strong QED has focussed on Schwinger
pair production (for references, see \cite{dunne-rev,ruffini}). However, the exact effective action determines the pair-production
rate through the vacuum persistence (twice of the imaginary part).

In the seminal paper \cite{schwinger} Schwinger introduced the proper-time integral to evaluate the determinant for the Dirac equation
and the Klein-Gordon equations in a constant electromagnetic field. In a constant magnetic field the evenly spaced energy spectrum
makes the zeta-function regularization available for the effective action \cite{blviwi}  (for review, see  \cite{eorbz}).
Also, the effective action can be obtained from various methods such as the worldline integral \cite{rescsc}, the Green's function
(resolvent method) \cite{chd,dunne-hall98,dunne-hall98-2} and the lightcone coordinate \cite{fried}.
The vacuum instability due to Schwinger pair production requires a state-of-the-art field theory for the effective action
in electric fields. In the St\"{u}ckelberg-Feynman picture for particle and antiparticle \cite{stuckelberg,feynman},
Nikishov showed that the second quantized field theory could not only explain Schwinger pair production in electric fields
in the time-dependent gauge \cite{nikishov} but also resolve the Klein paradox from barrier tunneling in the space-dependent gauge
\cite{nikishov70,nikishov79}, and Damour \cite{damour} and Hansen and Ravndal \cite{hansenravndal}
formulated the second quantized field theory in the space-dependent gauge
for electric fields. Schwinger pair production has also been studied in the second quantized field theory \cite{bmps,gagigo,gavgit,gasp,ggt,tanji,fgl,soldati}.

A temporal electromagnetic field or the time-dependent gauge for a constant electromagnetic field makes the out-vacuum in the future
differ from the in-vacuum in the past. There the in-out formalism based on the Schwinger variational principle \cite{schwinger51}
provides a natural framework for studying the effective action, which has been further developed by DeWitt \cite{dewitt,dewitt-book}.
In the in-out formalism the vacuum persistence amplitude (vacuum-to-vacuum transition amplitude or $S$-matrix) is the effective action
\begin{eqnarray}
e^{i \int d^3 x d t {\cal L}^{(1)} } = \langle {\rm out} \vert {\rm in} \rangle, \label{in-out act}
\end{eqnarray}
which takes the form \cite{dewitt,dewitt-book}
\begin{eqnarray}
\int d^3 x d t {\cal L}^{(1)} = \mp i \sum_{\bf K} \ln (\alpha_{\bf K}^*), \label{alp-act}
\end{eqnarray}
where and throughout the paper the upper (lower) sign is for spinor (scalar) QED, $\alpha_{\rm K}$ is Bogoliubov coefficient between the in-vacuum and the out-vacuum,
and ${\bf K}$ stands for all quantum numbers, such as the momentum, spin, and energy for the field.
The in-out formalism manifests the vacuum persistence relation
\begin{eqnarray}
2 {\rm Im} (\int d^3 x d t {\cal L}^{(1)}) = \mp  \sum_{\bf K} \ln (1 \mp |\beta_{\bf K}|^2), \label{vac per rel}
\end{eqnarray}
where the Bogoliubov relation holds
\begin{eqnarray}
|\alpha_{\bf K}|^2 \pm |\beta_{\bf K}|^2 = 1.
\end{eqnarray}
In QED it was pointed out in Ref. \cite{ahn} that the vacuum persistence amplitude could provide the effective action and
was shown in Ref. \cite{nikishov03} that the vacuum persistence amplitude recovered the effective action in a constant electromagnetic field.

Quite recently, Kim, Lee and Yoon have further developed the effective action in the in-out formalism in a temporary or a spatially
localized Sauter-type electric field \cite{kly08,kim09,kly10}. There, using the gamma-function regularization,
the vacuum persistence amplitudes from the Bogoliubov coefficients are shown to be the effective actions in these electric fields.
In the space-dependent gauge for electric fields the Bogoliubov coefficients from the second quantized field theory for
barrier tunneling \cite{nikishov70,nikishov79,damour,hansenravndal} are used to compute the effective actions \cite{kim09,kly10}.
The effective actions in electric fields in the in-out formalism can be readily extended to finite temperature ones \cite{kly10-2}.
Since finding QED effective actions in time-varying or spatially localized fields is quite nontrivial, the new method in the
in-out formalism may provide an alternative scheme to understand the vacuum structure under these field configurations.

The purpose of this paper is three-fold. First, we further develop the in-out formalism for QED effective actions in the space-dependent
gauge within the second quantized field theory. Second, we advance a method to find the QED effective actions in static magnetic fields
in the in-out formalism. Third, we show the electromagnetic duality of QED actions in a constant electric and a constant magnetic field
and also in a Sauter-type electric and a magnetic field. Though computationally efficient and powerful, the in-out formalism in
the space-dependent gauge requires some physical arguments how to select the in- and the out-vacua. We revisit the second quantized field theory
for barrier tunneling in electric fields in the space-dependent gauge and then find the effective action in a constant electromagnetic field.
We further consider a challenging problem whether the in-out formalism may be applicable to pure magnetic fields since bound states
of charged particles in magnetic fields, in particular, the Landau levels in a constant magnetic field make the in-out formalism
apparently suspicious. Contrary to a common belief, however, we argue that the inverse scattering matrix for charged particles in
pure magnetic fields may give the coefficient playing the same role as the Bogoliubov coefficient in the in-out formalism.
We use the inverse scattering matrix to find the effective actions in a constant and a Sauter-type magnetic field.

The underlying idea is that the exponentially decreasing and increasing solutions in each asymptotic region for charged particles
in pure magnetic fields can be written in the form of Jost functions for the scattering theory and the discrete spectrum of energy
is the consequence of imposing the boundary condition for normalizability and thereby on the Jost functions. This may be interpreted
as an on-shell condition for physically bound states. Unless the on-shell condition is imposed, in general one set of two independent
solutions with the required behavior in one asymptotic region can always be expressed as a linear combination of another set in the
other asymptotic region. This may be interpreted as an extension of scattering theory to bound states and to the relation among
solutions through the Jost functions, in which bound states occur when the scattering matrix has poles at the bound states \cite{taylor}.
This further extends the Jost functions for electric fields in the space-dependent gauge to bound systems for magnetic fields.
Remarkably the inverse scattering matrix, which is proportional to the amplitude of exponentially increasing branch, plays the
analogous role for the Bogoliubov coefficient and leads to the effective actions in pure magnetic fields.

The main difference from the previous works \cite{kly08,kim09,kly10} is that the Bogoliubov transformation and coefficients are found
from the Jost functions in the second quantized field theory and a method is proposed for effective actions for magnetic fields without
relying on the electromagnetic duality. The spin effect on the Bogoliubov coefficients is analyzed when both an electric field and
a magnetic field are present. Further we advance a method that generalizes the in-out formalism even to magnetic fields. The zeta-function
regularization does not easily apply to the spectrum in a general magnetic field, for instance, the Sauter-type magnetic field.
We observe that the inverse scattering matrix provides a method to find the effective action in analogy with the in-out formalism for
the tunneling picture in electric fields. We illustrate this method for a constant magnetic field and a spatially localized magnetic field
of Sauter-type.

We further show through one-loop effective actions in the in-out formalism that the electromagnetic duality holds between a constant electric
field and a constant magnetic field and also between an electric field and a magnetic field of Sauter-type. It does neither assume the presence
of electromagnetic field nor the imaginary discrete spectrum for the electric field. A common stratagem has been to show the duality
of the Heisenberg-Euler and Schwinger effective action in the constant electromagnetic field under the dual transformation of electric
field and magnetic field \cite{chopak,bcp}. In this paper we directly obtain the effective actions for magnetic fields of the same
form as electric fields using the inverse scattering matrix without imposing the on-shell condition. Then the effective actions,
unrenormalized or renormalized, exhibit the duality between the constant and the Sauter-type electric and magnetic fields.

The organization of this paper is as follows. In Sec. \ref{sec 2} we apply the second quantized field theory for barrier tunneling to QED,
find the Bogoliubov coefficients and obtain the effective action in an electric field parallel to a magnetic field.
In Sec. \ref{sec 3} we compare the in-out formalism in the time-dependent gauge and in the space-dependent gauge.
In Sec. \ref{sec 4} we advance a method to find effective action from the inverse scattering matrix in a pure magnetic field
and in Sec. \ref{sec 5} we apply the method to the Sauter-type magnetic field.
In Sec. \ref{sec 6} we show the duality of the one-loop effective actions in the constant and the Sauter-type electric and magnetic fields.

\section{In-Out Formalism in Space-Dependent Gauge} \label{sec 2}

We first find the effective actions (\ref{alp-act}) in spinor and scalar QED in a constant electromagnetic field in the space-dependent gauge
and then compare them with those in the time-dependent gauge \cite{kly08}. In the space-dependent (Coulomb) gauge for the electric field
the Dirac equation and the Klein-Gordon equation describe quantum tunneling phenomenon. We find the Bogoliubov coefficients in the
second quantized field theory for barrier tunneling.

A Lorentz frame may be found in which the constant electric and magnetic fields are parallel along the $z$-direction and
have the space-dependent gauge field
\begin{eqnarray}
A_{\mu} = (- E z, 0, - Bx, 0).
\end{eqnarray}
The spin-diagonal Fourier component of the squared Dirac equation [in units of $\hbar = c = 1$]
\begin{eqnarray}
\phi_{(r)}^{(\sigma)} (t, {\bf x}) = \int \frac{d\omega}{(2 \pi)} \frac{dk_y}{(2 \pi)}
e^{- i \omega t +i k_y y} \varphi_{(r)}^{(\sigma)} (x,z),
\end{eqnarray}
is given by (see Appendix \ref{dirac eq})
\begin{eqnarray}
\Bigl [\partial_x^2 - (k_y - q Bx)^2 + \partial_z^2 + (\omega + qEz)^2 - (m^2+ ir(qE) - 2 \sigma (qB)) \Bigr] \varphi_{(r)}^{(\sigma)} (x,z) = 0.
\end{eqnarray}
The eigenfunction for the motion transverse to the magnetic field is the harmonic wave function $\Phi_n (\sqrt{2/qB} (k_y - qBx))$ with spectrum $qB (2n+1)$ for $n = 0, 1, \cdots$, while the motion in the longitudinal direction is an inverted oscillator and describes a tunneling problem
\begin{eqnarray}
\Bigl[ \partial_z^2 + (\omega + qEz)^2 - (m^2 + qB (2n+1 - 2 \sigma) + ir(qE)) \Bigr] \varphi_{(r)n}^{(\sigma)} (z) = 0. \label{space sp-d four}
\end{eqnarray}

In the space-dependent gauge the in-vacuum is the Dirac sea in which virtual pairs annihilate themselves whereas the out-vacuum is the state
for particle-antiparticle pairs created from an electric field. In the St\"{u}ckelberg-Feynman picture \cite{stuckelberg,feynman},
a pair annihilation process for Eq. (\ref{space sp-d four}) is that a particle travels forward in time from one asymptotic
region $(z = - \infty)$ to the interaction region of electric field and then travels backward in time to another asymptotic
region $(z = \infty)$ while a pair production process is that an antiparticle travels backward in time from the region $z = \infty$,
interacts with the field, and then travels forward in time to the region $z = - \infty$. It was Nikishov \cite{nikishov70} who first
constructed the Green function in the space-dependent gauge for electric field, in which the space coordinate $z$ played the role of
time and wave functions were defined with respect to the flux $-i \partial_z$. Damour \cite{damour} and later Hansen
and Ravndal \cite{hansenravndal} also formluated the second quantized field theory for Schwinger pair production in space-dependent
gauge for electric fields. The second quantized field theory for the Bogoliubov transformations is illustrated in terms of Jost functions
in Appendix \ref{sec space-ga}. The wave functions for the in-vacuum are
\begin{eqnarray}
\Psi_{\rm (in)}^{(+)} &=& A_{\rm (in)} (\gamma^{\mu} P_{\mu} + m) D_{p} (- \zeta) \Xi_{(r)}^{(\sigma)}, \nonumber\\ 
\Psi_{\rm (in)}^{(-)} &=& B_{\rm (in)} (\gamma^{\mu} P_{\mu} + m) D_{-p-1} (-i \zeta) \Xi_{(r)}^{(\sigma)}, \label{lef sol}
\end{eqnarray}
where $D_p$ is the parabolic cylinder function \cite{gr-table}
\begin{eqnarray}
p = - \frac{1 + r}{2} + i \frac{m^2 + qB (2n+1 - 2 \sigma)}{2qE}, \quad \zeta = \sqrt{\frac{2}{qE}} e^{i \frac{\pi}{4}} (\omega + qEz).
\end{eqnarray}
Working in the Riemann sheet $ - \pi \leq {\rm arg} z < \pi$
\cite{riemann} and using the
asymptotic forms (\ref{par asy}), we find the wave functions for the out-vacuum
\begin{eqnarray}
\Psi_{\rm (out)}^{(+)} &=& A_{\rm (out)} (\gamma^{\mu} P_{\mu} + m) D_{-p-1} (i \zeta) \Xi_{(r)}^{(\sigma)}, \nonumber\\ 
\Psi_{\rm (out)}^{(-)} &=& B_{\rm (out)} (\gamma^{\mu} P_{\mu} + m) D_{p} (\zeta) \Xi_{(r)}^{(\sigma)}. \label{rig sol}
\end{eqnarray}
The normalization of the constants $A$ and $B$ will not be considered and a convention will be adopted that the spins of particle
and antiparticle are polarized along the magnetic field, i.e., the diagonal states of $\sigma^{12}$ but an average is taken
over the spin states of $\sigma^{03}$ for the electric field. The solutions (\ref{lef sol}) and (\ref{rig sol}) have the
WKB asymptotic form analogous to the Jost functions (\ref{jos asym2}).

Using the connection formula (\ref{asym 2}) and the relation (\ref{jos rel2}) between two sets of solutions
in terms of Jost functions, the particle wave function $\Psi_{\rm (in)}^{(+)}$ can be analytically continued to
\begin{eqnarray}
\Psi_{\rm (in)}^{(+)} = \sqrt{2 \pi} \frac{e^{- i(p+1) \frac{\pi}{2}}}{\Gamma (- p)} \Psi_{\rm (in)}^{(-)} + e^{-ip \pi} \Psi_{\rm (out)}^{(-)}, \label{bog1}
\end{eqnarray}
from which follow the Bogoliubov coefficients (\ref{space bog-fin})
\begin{eqnarray}
\alpha_{(r)n}^{(\sigma)} = \sqrt{2 \pi} \frac{e^{-i(2p^*+p+1) \frac{\pi}{2}}}{\Gamma (-p)}, \quad \beta_{(r)n}^{(\sigma)} = - e^{-ip^* \pi}.
\end{eqnarray}
The spin-averaged Bogoliubov coefficients over $(r)$ are given by \cite{popovmarinov}
\begin{eqnarray}
\alpha_{n}^{(\sigma)} = (\alpha_{(1)n}^{(\sigma)} \alpha_{(-1)n}^{(\sigma)})^{\frac{1}{2}}, \quad \beta_{n}^{(\sigma)} = (\beta_{(1)n}^{(\sigma)} \beta_{(-1)n}^{(\sigma)})^{\frac{1}{2}}.
\end{eqnarray}
The mean number of produced pairs for a given Landau level and spin state per unit volume and per unit time is
\begin{eqnarray}
{\cal N}_{n}^{(\sigma)} = |\beta_{n}^{(\sigma)}|^2 = e^{- \pi \frac{m^2 + qB (2n+1- 2 \sigma)}{qE}}, \label{mean num}
\end{eqnarray}
and the total mean numbers are the sum over the Landau levels and spin states
\begin{eqnarray}
{\cal N}_{\rm sp} &=& \frac{(qB)(qE)}{(2 \pi)^2} e^{- \frac{\pi m^2}{qE}} \coth(\frac{\pi B}{E}),\\ \nonumber\\
{\cal N}_{\rm sc} &=& \frac{(qB)(qE)}{2(2 \pi)^2} e^{- \frac{\pi m^2}{qE}} \frac{1}{\sinh(\frac{\pi B}{E})}. \label{tot mean num}
\end{eqnarray}
Here, the factor $(qB)/(2 \pi)$ accounts for Landau levels centered around $k_y = qBx$ while $(qE)/(2 \pi)$ is the number of
states from the wave packet around $\omega = - qEz$. Using the formulas (\ref{gamma form1}) for the gamma function, 
a direct calculation shows the Bogoliubov relation to hold
\begin{eqnarray}
|\alpha_{n}^{(\sigma)}|^2 = 1 \mp |\beta_{n}^{(\sigma)}|^2,
\end{eqnarray}
where the upper sign is for fermions and the lower sign is for bosons.

In the in-out formalism the effective action (\ref{alp-act}) is obtained by summing over $(n)$ and $(\sigma)$ for spinor QED
\begin{eqnarray}
{\cal L}^{(1)}_{\rm sp} = - i \frac{(qB)(qE)}{(2 \pi)^2} \sum_{n\sigma} \ln (\alpha^{(\sigma)*}_{n}) = - i \frac{(qB)(qE)}{2(2 \pi)^2} \sum_{n \sigma r} \ln (\alpha^{(\sigma)*}_{(r)n}).
\end{eqnarray}
and for scalar QED
\begin{eqnarray}
{\cal L}^{(1)}_{\rm sc} =  i \frac{(qB)(qE)}{(2 \pi)^2} \sum_{n} \ln (\alpha^{*}_{n}).
\end{eqnarray}
The gamma-function regularization \cite{kly08,kim09,kly10,kim10,kim11} leads to the unrenormalized effective action
\begin{eqnarray}
{\cal L}^{(1)}_{\rm sp} &=& i \frac{(qB)qE)}{2(2 \pi)^2} \sum_{n \sigma r} \ln \Gamma (-p^*) \nonumber\\
&=&  i \frac{(qB)(qE)}{2(2 \pi)^2} \sum_{n \sigma r} \int_{0}^{\infty} \frac{ds}{s} \frac{e^{p^* s}}{1 - e^{-s}}.
\end{eqnarray}
Here we have deleted all the terms independent of number of states and other divergent terms that are to be regulated away
through renormalization of the vacuum energy and the charge.

First, the renormalized effective action density for spinor QED, after summing over the spin states $(\sigma r)$ and the Landau levels, is given by
\begin{eqnarray}
{\cal L}^{(1)}_{\rm sp} &=&  \frac{(qB)(qE)}{2(2 \pi)^2} \int_{0}^{\infty} \frac{ds}{s} e^{-i \frac{m^2s}{2qE}} \Bigl[ \coth (\frac{s}{2}) \cot (\frac{Bs}{2E}) - \frac{1}{EB} \Bigl(\frac{4E^2}{s^2} - \frac{1}{3} (B^2 - E^2) \Bigr) \Bigr], \label{s sp act}
\end{eqnarray}
where the Schwinger prescription has been employed that subtracts the divergent terms in the proper-time integral \cite{schwinger}.
The first subtracted term corresponds to the vacuum energy renormalization and the second subtracted to the charge renormalization.
Either wick-rotating the proper time $s$ by $-is$ or doing a contour integral along an infinite quarterly circle in the fourth quadrant
as in Refs. \cite{kly08,kim09,kly10} and rescaling the proper time $s$ by $2qEs$, we recover the standard result
\begin{eqnarray}
{\cal L}^{(1)}_{\rm sp} =  - \frac{1}{2 (2 \pi)^2} {\cal P} \int_{0}^{\infty} \frac{ds}{s^3} e^{-m^2 s} \Bigl[\frac{(qEs) (qBs)}{\tan (qEs) \tanh (qBs)} - 1 - \frac{(qs)^2}{3} (B^2- E^2)\Bigr]. \label{sp act}
\end{eqnarray}
The sum of all residues from the principal value of Eq. (\ref{sp act}) or from the contour integral (\ref{s sp act}) leads to the vacuum persistence
\begin{eqnarray}
2 {\rm Im} ({\cal L}^{(1)}_{\rm sp}) = \frac{(qB)(qE)}{(2 \pi)^2} \sum_{k = 1}^{\infty} \frac{1}{k} e^{- \frac{k \pi m^2}{qE}} \coth(\frac{k \pi B}{E}). \label{sp vac per amp}
\end{eqnarray}
The vacuum persistence for spinor QED was obtained in Refs. \cite{heisenbergeuler,schwinger,bunkintugov,daughertylerche}.
Note that Eq. (\ref{sp vac per amp}) can be obtained from the relation (\ref{vac per rel}) by first expanding the logarithm
and then summing the mean number (\ref{mean num}) over the Landau levels and spin states, which is the consistency relation in the in-out formalism.

Second, the renormalized effective action in scalar QED is obtained by summing the Landau levels only
\begin{eqnarray}
{\cal L}^{(1)}_{\rm sc} =  - \frac{(qB)(qE)}{4(2 \pi)^2} \int_{0}^{\infty} \frac{ds}{s} e^{-i \frac{m^2s}{2qE}} \Bigl[ \frac{1}{\sinh (\frac{s}{2}) \sin (\frac{Bs}{2E})} - \frac{1}{EB} \Bigl(\frac{4E^2}{s^2} + \frac{1}{6} (B^2 - E^2) \Bigr) \Bigr]. \label{s sc act}
\end{eqnarray}
The vacuum persistence from all residues at simple poles
\begin{eqnarray}
2 {\rm Im} ({\cal L}^{(1)}_{\rm sc}) = \frac{(qB)(qE)}{2(2 \pi)^2} \sum_{k = 1}^{\infty} \frac{(-1)^{k+1}}{k} e^{-\frac{k \pi m^2}{qE}} \frac{1}{\sinh ( \frac{k \pi B}{E})}, \label{sc vac per amp}
\end{eqnarray}
is consistent with Refs. \cite{chopak,popov}. The vacuum persistence can also be obtained from the relation (\ref{vac per rel})
\begin{eqnarray}
2 {\rm Im} ({\cal L}^{(1)}_{\rm sc}) = \sum_{n = 0}^{\infty} \ln \Bigl( 1+ e^{- \pi \frac{m^2 + qB (2n+1)}{qE}} \Bigr),
\end{eqnarray}
by first expanding the logarithm and then summing over the Landau levels.
After doing a contour integral in the fourth quadrant and rescaling the proper time $s$ by $2qEs$, Eq. (\ref{s sc act}) recovers the standard result
\begin{eqnarray}
{\cal L}^{(1)}_{\rm sc} =  \frac{1}{4 (2 \pi)^2} {\cal P} \int_{0}^{\infty} \frac{ds}{s^3} e^{-m^2 s} \Bigl[\frac{(qEs) (qBs)}{\sin (qEs) \sinh (qBs)} -1 + \frac{(qs)^2}{6} (B^2- E^2)\Bigr]. \label{sc act}
\end{eqnarray}

\section{Comparison with Time-Dependent Gauge} \label{sec 3}

Now we find the effective actions (\ref{alp-act}) in the time-dependent gauge and compare them with those from the space-dependent
gauge in Sec. \ref{sec 2}. The time-dependent gauge field for the constant electromagnetic field is (see Appendix \ref{dirac eq})
\begin{eqnarray}
A_{\mu} = (0, 0, - Bx, Et).
\end{eqnarray}
After decomposing by Fourier modes and bispinors for spin states (\ref{bi sp})
\begin{eqnarray}
\phi_{(r)}^{(\sigma)} (t,{\bf x}) = \int \frac{dk_y}{(2 \pi)} \frac{dk_z}{(2 \pi)} e^{i (k_y y+ k_z z)} \varphi_{(r)}^{(\sigma)} (t,x),
\end{eqnarray}
and separating the Landau levels, the spin-diagonal component of the squared Dirac equation becomes a one-dimensional
scattering problem over the inverted oscillator potential
\begin{eqnarray}
\Bigl[ \partial_t^2 + (k_z + qEt)^2 + m^2 + qB (2n+1  - 2 \sigma)  + i r (q E) \Bigr] \varphi_{(r)n}^{(\sigma)} = 0.
\end{eqnarray}

The boundary condition is that the particle (antiparticle) should have a positive (negative) frequency with respect
to $i \partial_t$ in the past and in the future $(t = \mp \infty)$ as shown in Appendix \ref{sec time-ga}.
Then the wave functions for particle and antiparticle in the past are given by
\begin{eqnarray}
\Psi_{\rm (in)}^{(+)} &=& A_{\rm (in)} (\gamma^{\mu} P_{\mu} + m) D_{-p-1} (i \eta) \Xi_{(r)}^{(\sigma)}, \nonumber\\ \Psi_{\rm (in)}^{(-)} &=& B_{\rm (in)} (\gamma^{\mu} P_{\mu} + m) D_{p} (- \eta) \Xi_{(r)}^{(\sigma)},
\end{eqnarray}
where
\begin{eqnarray}
\eta = \sqrt{\frac{2}{qE}} e^{i \frac{\pi}{4}} (k_z + qEt), \quad p = - \frac{1-r}{2} - i \frac{m^2 + qB (2n+1 - 2 \sigma)}{2qE}.
\end{eqnarray}
Similarly, the wave functions for particle and antiparticle in the future are
\begin{eqnarray}
\Psi_{\rm (out)}^{(+)} &=& A_{\rm (out)} (\gamma^{\mu} P_{\mu} + m) D_{p} (\eta) \Xi_{(r)}^{(\sigma)}, \nonumber\\ \Psi_{\rm (out)}^{(-)} &=& B_{\rm (out)} (\gamma^{\mu} P_{\mu} + m) D_{-p-1} (-i \eta) \Xi_{(r)}^{(\sigma)}.
\end{eqnarray}
The constants $A$ and $B$ are determined by the normalization condition in the past and in the future, respectively, and
may be found in Ref. \cite{soldati}.

Applying the connection formula (\ref{asym 3}) to the particle wave function in the past,
\begin{eqnarray}
\Psi_{\rm (in)}^{(+)} = \frac{\sqrt{2 \pi}}{\Gamma(p+1)} e^{-i p \frac{\pi}{2}} \Psi_{\rm (out)}^{(+)} + e^{-i (p+1) \pi} \Psi_{\rm (out)}^{(-)}, \label{bog2}
\end{eqnarray}
the spin-averaged Bogoliubov coefficients
\begin{eqnarray}
\alpha_{n}^{(\sigma)} = (\alpha_{(1)n}^{(\sigma)} \alpha_{(-1)n}^{(\sigma)})^{\frac{1}{2}}, \quad
\beta_{n}^{(\sigma)} = (\beta_{(1)n}^{(\sigma)} \beta_{(-1)n}^{(\sigma)})^{\frac{1}{2}},
\end{eqnarray}
can be found from Eqs. (\ref{time bog1}) and (\ref{time bog2}), which are
\begin{eqnarray}
\alpha_{(r)n}^{(\sigma)} = \frac{\sqrt{2 \pi}}{\Gamma(p+1)} e^{-ip \frac{\pi}{2}}, \quad \beta_{(r)n}^{(\sigma)} = e^{i (p^*+1) \pi}. 
\label{time bog coef2}
\end{eqnarray}
The formulas (\ref{gamma form1}) of the gamma function leads to the Bogoliubov relation
\begin{eqnarray}
|\alpha_{n}^{(\sigma)}|^2 \pm |\beta_{n}^{(\sigma)}|^2 =1.
\end{eqnarray}
Following the procedure in Sec. \ref{sec 2}, from the Bogoliubov coefficient (\ref{time bog coef2}) we obtain the same  
effective actions (\ref{s sp act}) for spinor QED and (\ref{s sc act}) for scalar QED
in the in-out formalism.

\section{Constant Magnetic Field} \label{sec 4}

We turn to effective actions in pure magnetic fields, the main issue of this paper, and illustrate how the in-out formalism
may be generalized to such a bounded system for charged particles. In a constant magnetic field the transverse motion of a
charged particle is confined to Landau levels and in a general configuration has still a discrete spectrum of energy.
The vacuum defined by the lowest Landau level is stable and thus no pair is produced from pure magnetic fields.
In fact, a charged particle in pure magnetic fields has infinite instanton action and thus the probability for the vacuum
to decay and for pair production via instanton is essentially zero \cite{kimpage02,kimpage06}.

The spin-diagonal component of the squared Dirac or Klein-Gordon equation
\begin{eqnarray}
\Bigl [\partial_x^2 - (k_y - q Bx)^2 + \omega^2 - m^2 - k_z^2 + 2\sigma (qB) \Bigr] \varphi_{(\sigma)} (x) = 0, \label{mag eq}
\end{eqnarray}
shows that the bound states are harmonic wave functions with the energy $\epsilon = \omega^2 - m^2 - k_z^2 + 2\sigma (qB)$ corresponding
to the Landau levels $\epsilon = qB (2n+1)$. The general solutions for (\ref{mag eq}) are
\begin{eqnarray}
D_p (\xi), \quad D_p(- \xi), \quad D_{-p-1} (i \xi), \quad D_{-p-1} (-i \xi), \label{con mag sol}
\end{eqnarray}
where
\begin{eqnarray}
\xi = \sqrt{\frac{2}{qB}} (k_y - qBx), \quad p = - \frac{1- 2\sigma}{2} + \frac{\omega^2 - m^2 - k_z^2}{2qB}.
\end{eqnarray}
The exponentially decreasing solutions are $D_p (-\xi)$ at $x = \infty$ and $D_{p} (\xi)$ at $x = - \infty$ while the exponentially
increasing solutions are $D_{-p-1} (-i\xi)$ at $x = \infty$ and $D_{-p-1} (i\xi)$ at $x = - \infty$. As shown in Appendix \ref{sec mag-ga},
these functions can be used as the Jost functions for the bounded system as a generalization of scattering theory.
In fact, the connection formula (\ref{asym 3}) connects the bounded solution at $x = \infty$
\begin{eqnarray}
D_{p} (- \xi) = \sqrt{2 \pi} \frac{e^{i(p+1) \frac{\pi}{2}}}{\Gamma (-p)} D_{-p-1} (i \xi) + e^{ip \pi} D_{p} (\xi), 
\end{eqnarray}
 in terms of the Jost functions through Eq. (\ref{jos rel3}).
The relation is reminiscent of Eqs. (\ref{bog1}) and (\ref{bog2}) that lead to the Bogoliubov transformation and
coefficients for the constant electric field together with the constant magnetic field.
We may introduce the inverse scattering matrix (\ref{in sc mat}), which is the ratio of the amplitude for exponentially increasing
part to the amplitude for exponentially decreasing part
\begin{eqnarray}
{\cal M}_p = \sqrt{2 \pi} \frac{e^{-i(p-1) \frac{\pi}{2}}}{\Gamma (-p)}. \label{sc mat}
\end{eqnarray}
Note that the inverse scattering matrix (\ref{sc mat}) is indeed the inverse of the scattering matrix in scattering theory, in which
the scattering matrix is the ratio of the amplitude for exponentially decreasing part to the amplitude for the exponentially
increasing part \cite{taylor}.

The inverse scattering matrix now carries the information about the potential and quantum states. For instance, the condition for bound states is
\begin{eqnarray}
{\cal M}_p = 0, \quad p = n, \quad (n = 0, 1, \cdots). \label{quan1}
\end{eqnarray}
The simple poles for the scattering matrix at physically bound states \cite{taylor} now become the simple zeros of $1/\Gamma (-p)$
for the inverse scattering matrix, in which $D_n(\xi)$ is the harmonic wave function up to a normalization constant with parity $e^{ip \pi}$.
That is, the nonnegative integer $p=n$ is the on-shell condition for Landau levels. Wick-rotating the time as $t = -i\tilde{t}$
and the frequency as $\omega = i \tilde{\omega}$, we observe that the inverse scattering matrix provides the effective action
in analogy with the in-out formalism for electric fields
\begin{eqnarray}
{\cal L}^{(1)} = \pm \frac{qB}{(2 \pi)} \sum_{\sigma} \int \frac{d \tilde{\omega}}{(2 \pi)} \frac{dk_z}{(2 \pi)} \ln ({\cal M}_p^*).
\end{eqnarray}
where the upper sign is for spinor QED and the lower sign is for scalar QED.
Using the gamma-function regularization, summing over the spin states and carrying out the integration, we obtain the effective action for spinor QED
\begin{eqnarray}
{\cal L}^{(1)}_{\rm sp} =  - \frac{(qB)^2}{(2\pi)^2}\int_{0}^{\infty} \frac{ds}{s^2} e^{- \frac{m^2s}{2qB}} \Bigl(\coth(\frac{s}{2}) - \frac{2}{s} - \frac{s}{6} \Bigr),
\end{eqnarray}
and for scalar QED
\begin{eqnarray}
{\cal L}^{(1)}_{\rm sc} =  \frac{(qB)^2}{2(2\pi)^2}\int_{0}^{\infty} \frac{ds}{s^2} e^{- \frac{m^2s}{2qB}} \Bigl( \frac{1}{\sinh(\frac{s}{2})} - \frac{2}{s} + \frac{s}{12} \Bigr).
\end{eqnarray}
Thus the inverse scattering matrix method recovers the standard result (\ref{sp act}) and (\ref{sc act}) for the constant magnetic
field in the limit of $E = 0$, as expected.

A passing remark is that the inverse scattering matrix is real and therefore the effective action does not have an imaginary part
and does not lead the vacuum decay due to pair production. It remains open to show the connection between the on-shell approach
using the Green's function from bounded states \cite{chd,dunne-hall98-2} and the off-shell approach using the inverse scattering matrix
in the in-out formalism, which is beyond the scope of this paper. Though DeWitt proved the equivalence between
two approaches in a general context of field theory, it would be interesting to show explicitly the equivalence using
the Jost functions (\ref{bog coef1}) and (\ref{bog coef2}) as the Bogoliubov coefficients for electric fields and
the Jost functions (\ref{bog coef3}) for magnetic fields since they are the Wronskian of two independent solutions.

\section{Spatially Localized Magnetic Field} \label{sec 5}

We now consider a spatially localized field $B(x) = B \, {\rm sech}^2(x/L)$ along the $z$-direction with the space-dependent gauge field
\begin{eqnarray}
A_{\mu} = (0, 0, - BL \tanh (\frac{x}{L}), 0).
\end{eqnarray}
The spin-diagonal Fourier component of squared Dirac equation becomes
\begin{eqnarray}
\Bigl [\partial_x^2 - (k_y - q BL \tanh (\frac{x}{L}))^2 + \omega^2 - m^2 - k_z^2 +2 \sigma qB {\rm sech}^2 (x/L) \Bigr] \varphi_{(\sigma)} (x) = 0. \label{mag eq2}
\end{eqnarray}
The motion (\ref{mag eq2}) is bounded at $x = \pm \infty$, so the momentum $P_1$ takes imaginary values, $P_{1(\pm)} = i\Pi_{1(\pm)}$,
\begin{eqnarray}
\Pi_{1(\pm)} = \sqrt{(k_y \mp q BL)^2 - (\omega^2 - m^2 - k_z^2)}.
\end{eqnarray}
Then the solution may be found in terms of the hypergeometric function as
\begin{eqnarray}
\varphi_{(\sigma)} (x) = \xi^{\frac{L}{2} \Pi_{1(+)}} (1 - \xi)^{\frac{1-2\sigma}{2} + \lambda_{\sigma}} F(a, b; c; \xi),
\end{eqnarray}
where
\begin{eqnarray}
\xi = - e^{- 2 \frac{x}{L}}, \quad \lambda_{\sigma} = \sqrt{(qBL^2)^2 + \Bigl(\frac{1-2|\sigma|}{2} \Bigr)^2},
\end{eqnarray}
and
\begin{eqnarray}
a &=& \frac{1-2\sigma}{2} + \frac{1}{2} (L\Pi_{1(+)}+ L\Pi_{1(-)} +2\lambda_{\sigma}) := \frac{1-2\sigma}{2} + \frac{\Omega_{(+)}}{2}, \nonumber\\
b &=& \frac{1-2\sigma}{2} + \frac{1}{2} (L\Pi_{1(+)} - L\Pi_{1(-)} +2\lambda_{\sigma}) := \frac{1-2\sigma}{2} + \frac{\Delta_{(+)}}{2}, \nonumber\\
c &=& 1 + L \Pi_{1(+)}.
\end{eqnarray}

The solution is bounded at $x = \infty \, (\xi = 0)$ since $\Pi_{1(+)}$ is positive. In the opposite limit $x = - \infty \, (\xi = - \infty)$,
using the connection formula (\ref{hyper con}), we find the asymptotic form for the solution
\begin{eqnarray}
\varphi_{(\sigma)} = (-1)^{\frac{L}{2} \Pi_{1(+)}} \Bigl[
(- \xi)^{-\frac{L}{2} \Pi_{1(-)}} \frac{\Gamma (c) \Gamma(b -a)}{\Gamma(b) \Gamma(c-a)} + (- \xi)^{\frac{L}{2} \Pi_{1(+)}} \frac{\Gamma (c) \Gamma(a -b)}{\Gamma(a) \Gamma(c-b)} \Bigr]. \label{mag sa con}
\end{eqnarray}
The first term exponentially decreases while the second term increases. As $a > b > 0$ and $c > b$, the condition for bound states is
that $\Gamma (c-b)$ should be singular, which leads to
\begin{eqnarray}
c - b = \frac{L}{2} (\Pi_{1(+)} + \Pi_{1(-)}) - \lambda_{\sigma} + \frac{1+ 2 \sigma}{2} = - n, \quad (n =0, 1, \cdots). \label{quan2}
\end{eqnarray}
There is a finite number of discrete spectrum
\begin{eqnarray}
\omega^2 = m^2 + k_z^2 + (k_y + qBL)^2 - \Bigl(
\frac{2 \lambda_{\sigma} - 2n -1 - 2 \sigma}{2L} + \frac{2k_y qBL^2}{2 \lambda_{\sigma} - 2n -1 - 2\sigma} \Bigr)^2, \label{spec2}
\end{eqnarray}
with $n + (1+ 2\sigma)/2 <\lambda_{\sigma}$ from Eq. (\ref{quan2}). In the limit of $qBL \gg |\omega|$, the discrete spectrum (\ref{quan2})
approaches the Landau levels (\ref{quan1}). The zeta-regularization to the spectrum (\ref{spec2})
becomes nontrivial in finding the effective action.

Now we apply the method of the inverse scattering matrix in Sec. \ref{sec 4}. The asymptotic solutions (\ref{jos fn3}) are
$\xi^{L\Pi_{1(+)}/2}$ at $x = \infty$ and $(-\xi)^{-L\Pi_{1(-)}/2}$ at $x = - \infty$, so Eq. (\ref{mag sa con}) connects
the solutions in terms of the Jost functions (\ref{jos fn3}). Then the inverse scattering matrix is
\begin{eqnarray}
 {\cal M} = \frac{\Gamma (b) \Gamma(c-a)}{\Gamma(a) \Gamma (c-b)}, \label{sc-Bsa}
\end{eqnarray}
where $\Gamma (a-b)/\Gamma(b-a)$ that depend only $\Pi_{1(-)}$ can be gauged away by choosing $A_2 = BL (\tanh (x/L) + 1)$
and will not be included hereafter. In the limit of $qBL \gg |\omega|$, the inverse scattering matrix reduces to
\begin{eqnarray}
{\cal M} = \Bigl( \frac{\Gamma (qBL^2)\Gamma (-qBL^2)}{\Gamma(2qBL^2)} \Bigr) \frac{1}{\Gamma(-p)},
\end{eqnarray}
with $p$ from Eq. (\ref{quan1}). The terms in the bracket are independent of the number of states and are to be regulated away
through renormalization of the effective action. Thus the last factor gives the effective action in a constant magnetic field
 in this limit as in Sec. \ref{sec 4}. Applying the identity  of the gamma function (\ref{gamma form2}) to negative values of
\begin{eqnarray}
c-a &=& \frac{1+ 2 \sigma}{2} + \frac{1}{2} (L\Pi_{1(+)} - L\Pi_{1(-)} - 2\lambda_{\sigma}) := \frac{1+2\sigma}{2} + \frac{\Delta_{(-)}}{2}, \nonumber\\
c-b &=& \frac{1+2\sigma}{2} + \frac{1}{2} (L\Pi_{1(+)} + L\Pi_{1(-)} - 2\lambda_{\sigma}) := \frac{1+ 2\sigma}{2} + \frac{\Omega_{(-)}}{2},
\end{eqnarray}
we may write the inverse scattering matrix as
\begin{eqnarray}
 {\cal M} = \frac{\Gamma (\frac{1-2\sigma}{2} + \frac{\Delta_{(+)}}{2}) \Gamma(\frac{1-2\sigma}{2} - \frac{\Omega_{(-)}}{2})}{\Gamma(\frac{1-2\sigma}{2} - \frac{\Delta_{(-)}}{2}) \Gamma (\frac{1-2\sigma}{2} + \frac{\Omega_{(+)}}{2})}. \label{sc-Bsa2}
\end{eqnarray}
Here we have deleted again the overall factor that is to be regulated away through the renormalization procedure.

Finally, the gamma-function regularization leads to the renormalized effective action in spinor QED
\begin{eqnarray}
{\cal L}^{(1)}_{\rm sp} &=& - \int \frac{d \tilde{\omega}}{(2\pi)} \frac{d^2 {\bf k}_{\perp}}{(2 \pi)^2}
\int_{0}^{\infty} \frac{ds}{s}  \Bigl(e^{- \frac{\Omega_{(+)}}{2}s} + e^{\frac{\Delta_{(-)}}{2}s} - e^{\frac{\Omega_{(-)}}{2}s} - e^{- \frac{\Delta_{(+)}}{2}s} \Bigr) \nonumber\\
&& \qquad \qquad \qquad \qquad \qquad \times \Bigl( \coth (\frac{s}{2}) - \frac{2}{s} - \frac{s}{6} \Bigr), \label{sa ma act}
\end{eqnarray}
and in scalar QED
\begin{eqnarray}
{\cal L}^{(1)}_{\rm sc} &=& \frac{1}{2} \int \frac{d \tilde{\omega}}{(2\pi)} \frac{d^2 {\bf k}_{\perp}}{(2 \pi)^2}
\int_{0}^{\infty} \frac{ds}{s}  \Bigl(e^{- \frac{\Omega_{(+)}}{2}s} + e^{\frac{\Delta_{(-)}}{2}s} - e^{\frac{\Omega_{(-)}}{2}s} - e^{- \frac{\Delta_{(+)}}{2}s} \Bigr)
\nonumber\\
&& \qquad \qquad \qquad \qquad \qquad \times \Bigl( \frac{1}{\sinh(\frac{s}{2})} - \frac{2}{s} + \frac{s}{12}\Bigr).
\end{eqnarray}
Here $\Omega_{(-)} <0 $ and $\Delta_{(-)} < 0$ and we have taken the Wick-rotation $t = - i \tilde{t}$ and $\omega = i \tilde{\omega}$ and used the Schwinger prescription for renormalization.
It would be interesting to compare the effective action (\ref{sa ma act}) with that in Ref. \cite{dunne-hall98-2} from the Green's function
(resolvent method).

We briefly explain a scheme to find approximately the effective action for a general configuration of magnetic field,
when solutions are not known. We assume a spatially localized field $B(x)$ along the $z$-direction and the gauge field $A_3$
such that $B(x) = \partial_x A_3(x)$.
Then the spin-diagonal component of squared Dirac equation will become
\begin{eqnarray}
\Bigl[ \partial_x^2  - \Pi_{1}^2 (x) \Bigr] \varphi_{(\sigma)} (x) = 0. \label{mag eq3}
\end{eqnarray}
where
\begin{eqnarray}
\Pi_{1}^2 (x) =  (k_y - q A_3(x))^2 - \omega^2 + m^2 + k_z^2 - 2 \sigma qB (x).
\end{eqnarray}
The uniform semiclassical approximation for electric fields \cite{dunne-hall98,kly10} suggests transforming (\ref{mag eq3}) into the form
\begin{eqnarray}
\Bigl[\partial_{\xi}^2 - \xi^2 + \frac{{\cal S}_{(\sigma)}}{\pi} + \frac{1}{(\partial_{x} \xi)^{3/2}} \partial_x^2 (\frac{1}{\sqrt{\partial_x \xi}}) \Bigr] w_{(\sigma)} (\xi) = 0, \label{b app eq}
\end{eqnarray}
where
\begin{eqnarray}
w_{(\sigma)} (\xi) = \sqrt{\partial_x \xi} \varphi_{(\sigma)} (x), \quad \Bigl(\xi^2 - \frac{{\cal S}_{(\sigma)}}{\pi} \Bigr) (\partial_x \xi)^2 = \Pi_{1}^2. \label{b ins}
\end{eqnarray}
The charged particle undergoes a periodic motion in the region $\Pi_{1}^2 \leq 0$, so the integration of Eq. (\ref{b ins}) over one period determines
the action \begin{eqnarray}
{\cal S}_{(\sigma)} = \oint \sqrt{- \Pi_{1}^2 (x)}dx.
\end{eqnarray}
Thus, in the approximation of neglecting the last term, Eq. (\ref{b app eq}) has the same form as Eq. (\ref{mag eq}) for the constant
magnetic field and the approximate solutions are $D_p (\sqrt{2} \xi)$, $D_p(- \sqrt{2}\xi)$, $D_{-p-1} (i \sqrt{2} \xi)$,
and $D_{-p-1} (-i \sqrt{2}\xi)$ with
\begin{eqnarray}
p = - \frac{1}{2} + \frac{{\cal S}_{(\sigma)}}{\pi}. \label{b app p}
\end{eqnarray}
Then the inverse scattering matrix is given by Eq. (\ref{sc mat}) with $p$ in Eq. (\ref{b app p}).
As $B(x) \leftrightarrow - B(x)$ under $k_y \leftrightarrow - k_y$ and $\sigma \leftrightarrow - \sigma$, we find the
unrenormalized effective action in a symmetric form
\begin{eqnarray}
{\cal L}^{(1)} &=& \mp \frac{1}{2} \sum_{\sigma} \int \frac{d \tilde{\omega}}{(2\pi)} \frac{d^2 {\bf k}_{\perp}}{(2 \pi)^2} \Bigl[ \ln \Gamma (- p (B) ) + \ln \Gamma (- p (-B) ) \Bigr],
\end{eqnarray}
where the upper sign is for spinor QED and the lower sign is for scalar QED. The renormalization procedure is the same as the constant meagnetic field.

\section{Electromagnetic Duality of QED Actions} \label{sec 6}

To show the electromagnetic duality, we recapitulate the main results of Refs. \cite{kim09,kly10} for the constant electric
field and the Sauter-type electric field in the space-dependent gauge. In the space-dependent gauge $({\bf E} = - \nabla A_0)$ 
for the pure electric field along the $z$-direction
\begin{eqnarray}
A_{\mu} = (A_0(z), 0, 0, 0),
\end{eqnarray}
the spin-diagonal Fourier component of the squared Dirac equation is
\begin{eqnarray}
\Bigl [\partial_z^2 + (\omega - qA_0(z))^2 - (m^2 + {\bf k}_{\perp}^2 + ir q E(z)) \Bigr] \varphi_{(r)}(z) = 0.
\end{eqnarray}

First, in the constant electric field with $A_0 (z) = - Ez$, the particle
and antiparticle wave functions for the in-vacuum are given by Eq. (\ref{lef sol}) and
for the out-vacuum by Eq. (\ref{rig sol}) with 
\begin{eqnarray}
p = - \frac{1 + r}{2} + i \frac{m^2 + {\bf k}_{\perp}^2}{2qE}.
\end{eqnarray}
Therefore, according to Sec. \ref{sec 2}, the Bogoliubov coefficients in the in-out formalism
\begin{eqnarray}
\alpha_{(r)} = \sqrt{2 \pi} \frac{e^{-i(2p^*+p+1) \frac{\pi}{2}}}{\Gamma (-p)}, \quad \beta_{(r)} = - e^{-ip^* \pi},
\end{eqnarray}
leads to the unrenormalized effective action
\begin{eqnarray}
{\cal L}^{(1)}_{\rm sp} &=& i \frac{qE}{2(2 \pi)} \sum_{\sigma r} \int \frac{d^2 {\bf k}_{\perp}^2}{(2\pi)^2}
\int_{0}^{\infty} \frac{ds}{s} \frac{e^{-p^*s}}{1 - e^{-s}} \nonumber\\
&=& - \frac{(qE)^2}{(2 \pi)^2} \int_{0}^{\infty} \frac{ds}{s^2} \cot(\frac{s}{2}) e^{- \frac{m^2 s}{2qE}}.
\end{eqnarray}
The renormalized effective action is obtained by replacing $\cot(s/2)$ by $\cot(s/2)- s/2 + s/6$.
The $\Gamma (-p^*)$ in $\alpha_{(r)}^*$ for the effective action is the same as that from the inverse scattering matrix (\ref{sc mat})
provided that $E = i B$ and $\omega = i \tilde{\omega}$. This implies that the unrenormalized and the renormalized
effective actions for the electric field are dual to those for the magnetic field. The duality also holds for scalar QED.

Second, in the Sauter-type electric field $E(z) = E \, {\rm sech}^2(z/L)$ the space-dependent gauge field is
\begin{eqnarray}
A_{\mu} = (- EL \tanh (\frac{z}{L}), 0, 0, 0).
\end{eqnarray}
The effective action in the electric field $E(z) = E \, {\rm sech}^2(t/T)$ was studied in Refs. \cite{kly08,dunne-hall98}
and in electric field $E(z) = E \, {\rm sech}^2(z/L)$ in Ref. \cite{kly10}.
The spin-diagonal Fourier component of squared Dirac equation takes the form
\begin{eqnarray}
\Bigl [\partial_z^2 + (\omega + q EL \tanh (\frac{z}{L}))^2 - i r (q E) {\rm sech}^2 (z/L)-
(m^2 + k_x^2 + k_z^2) \Bigr] \varphi_{(\sigma)} (z) = 0, \label{el eq2}
\end{eqnarray}
which has the asymptotic longitudinal momentum
\begin{eqnarray}
P_{3(\pm)} = \sqrt{(\omega \mp qEL)^2 - (m^2+ k_x^2 + k_y^2)}.
\end{eqnarray}
The Bogoliubov coefficient is given by \cite{kly10}
\begin{eqnarray}
\alpha_{(r)} = \frac{\Gamma (\tilde{b}) \Gamma(\tilde{c}-\tilde{a})}{\Gamma(\tilde{a}) \Gamma (\tilde{c}-\tilde{b})}, \label{bog-Esa}
\end{eqnarray}
where
\begin{eqnarray}
\tilde{a} &=& \frac{1+r}{2} -  \frac{i}{2} (L P_{3(+)}+ LP_{3(-)} - 2\lambda_{r}) := \frac{1+r}{2} -  i \frac{\tilde{\Omega}_{(-)}}{2}, \nonumber\\
\tilde{b} &=& \frac{1+r}{2} - \frac{i}{2} (LP_{3(+)} - LP_{3(-)} - 2\lambda_{r}) := \frac{1+r}{2} - i \frac{\tilde{\Delta}_{(-)}}{2}, \nonumber\\
\tilde{c} &=& 1 -i L P_{3(+)}.
\end{eqnarray}
and
\begin{eqnarray}
\tilde{c} - \tilde{a} &=& \frac{1-r}{2} - \frac{i}{2} (L P_{3(+)} - LP_{3(-)} + 2\lambda_{r}) := \frac{1+r}{2} -  i \frac{\tilde{\Delta}_{(+)}}{2}, \nonumber\\
\tilde{c} - \tilde{b} &=& \frac{1-r}{2} - \frac{i}{2} (LP_{3(+)} + LP_{3(-)} + 2\lambda_{r}) := \frac{1+r}{2} - i \frac{\tilde{\Omega}_{(+)}}{2}.
\end{eqnarray}
Under $E = iB$ and $\omega=i \tilde{\omega}$ and the interchange of $\tilde{\omega} \leftrightarrow -k_y$
and $k_x \leftrightarrow k_z$, we have $P_{3 (\pm)} = - i \Pi_{1(\pm)}$, $\lambda_{r} = -i \lambda_{\sigma}$ \cite{riemann} 
and thus $\tilde{\Omega}_{(\pm)} = - i \Omega_{(\pm)}$, 
$\tilde{\Delta}_{(\pm)} = - i \Delta_{(\pm)}$  and the coefficient $\alpha_{(r)}^*$ has the same form as 
the inverse scattering matrix (\ref{sc-Bsa}).
This shows that the Bogoliubov coefficient (\ref{bog-Esa}) can be analytically continued to the inverse scattering
matrix (\ref{sc-Bsa}) for the magnetic field and vice versa and that the unrenormalized and the renormalized
effective actions in the Sauter-type electric field are dual to those in the Sauter-type magnetic field.

In summary, we have shown in the in-out formalism that QED effective actions are dual not only for constant electric
and magnetic fields but also for the Sauter-type electric and magnetic fields. The duality of QED effective actions is also
consistent with the duality under $E = i B$ of the convergent series for the Heisenberg-Euler and Schwinger effective actions
(\ref{sp act}) and (\ref{sc act}) in constant electric and magnetic fields \cite{bcp}.

\section{Conclusion} \label{conclusion}

In this paper we have generalized the in-out formalism for QED effective actions to electromagnetic fields in the space-dependent gauge
and proposed a method to find the one-loop effective action from the inverse scattering matrix for magnetic fields.
In the space-dependent gauge for electric fields, the spin-diagonal Fourier component for the Dirac equation encounters
a potential barrier, in which the Dirac sea is lowered by the electrostatic potential, and by which spin-1/2 fermions suffer from the Klein paradox.
In contrast to the time-dependent gauge, in which the in- and the out-vacua are defined
in the past and in the future, in the second quantized field theory for barrier tunneling, the in-vacuum is the Dirac sea in which
pairs are annihilated and the out-vacuum is the state for particle-antiparticle pairs created from electric fields. We have applied
the second quantized field theory to a constant electromagnetic field in the space-dependent gauge and found 
the effective action from the Bogoliubov coefficients, which are related to the Jost functions illustrated in Appendix \ref{sec space-ga}.

We have further developed QED effective actions \cite{kly08,kim09,kly10} 
in the in-out formalism based on the Schwinger variational principle.
In the in-out formalism the effective action is the vacuum persistence amplitude or the $S$-matrix between the out-vacuum and
the in-vacuum, which is determined by the Bogoliubov coefficient. Recently the vacuum persistence amplitude has been used to find
QED effective actions in a constant electric field and a Sauter-type electric field in the time-dependent gauge \cite{kly08}
and a Sauter-type electric field in the space-dependent gauge \cite{kly10}. As QED effective actions in time-varying or
spatially localized fields are nontrivial, the new method in the in-out formalism may provide an alternative scheme to understand
the vacuum structure.

Remarkably the in-out formalism may have an extension to magnetic fields when the inverse scattering matrix
is used for the Bogoliubov coefficient. As shown in Sec. \ref{sec 4}, \ref{sec 5} and Appendix \ref{sec mag-ga}, the Jost functions
for off-shell solutions give the inverse scattering matrix in analogy with the Jost functions for the Bogoliubov coefficients in
electric fields in the space-dependent gauge. The inverse scattering matrix is the ratio of the amplitude of exponentially increasing
part to the amplitude of the exponentially decreasing part. The bound states occur at zeros of the inverse scattering matrix,
which correspond to poles of the scattering matrix. We have illustrated the method by computing the effective actions in a constant magnetic
field and a localized magnetic field of Sauter-type.

The effective actions in the in-out formalism are logarithms of the Bogoliubov coefficient
for electric fields and of the inverse scattering matrix for magnetic fields in the second quantized field theory in the space-dependent gauge.
In fact the Bogoliubov coefficient is determined by the Jost functions for tunneling solutions through barrier in electric fields while the
inverse scattering matrix also is determined by the Jost functions for off-shell solutions for magnetic fields, both of which are
Wronskian for the solutions with required asymptotic behaviors in two asymptotic regions. For this reason these coefficients are
analytically continued to each other under the electromagnetic duality and QED effective actions, renormalized or unrenormalized,
are dual to electric and magnetic fields of the same profile. We have explicitly showed the electromagnetic duality of QED effective
actions in constant and Sauter-type electric and magnetic fields.

\acknowledgments

The author would like to thank Don N. Page for early collaboration on Schwinger pair production in the space-dependent gauge,
Hyun Kyu Lee and Yongsung Yoon for collaboration on QED effective action, W-Y. Pauchy Hwang for helpful discussion on the in-out
formalism and Wei-Tou Ni
for useful discussion on effective action in localized magnetic fields and birefringence.
He also appreciates the warm hospitality at Institute of Astrophysics, National Taiwan University and thanks
Misao Sasaki for the warm hospitality at Yukawa Institute for Theoretical Physics, Kyoto University, where this paper was completed, and
Christian Schubert for the warm hospitality at Instituto de F\'{i}sica y Matem\'{a}ticas, Universidad Michoacana
de San Nicol\'{a}s de Hidalgo, where this paper was revised. This work was supported in part by Basic Science Research Program through
the National Research Foundation of Korea (NRF) funded by the Ministry of Education, Science and Technology (2011-0002-520)
and in part by National Science Council Grant (NSC 100-2811-M-002-012) by the Taiwan Government.

\appendix

\section{Dirac Equation in Space-Dependent Gauge} \label{dirac eq}

We use the notation (for instance, see Ref. \cite{greiner-book}) for the metric tensor $g_{\mu \nu} = g^{\mu \nu}=  (+,-,-,-)$,
the contravariant vector $x^{\mu} = (t, {\bf x})$, the covariant gauge field $A_{\mu} = (A_0, -{\bf A})$,
and the covariant derivative for the minimal coupling
\begin{eqnarray}
{\cal P}_{\mu} = i \partial_{\mu} - q A_{\mu} = i D_{\mu} = (i \partial_t - qA_0, i \nabla + q {\bf A}).
\end{eqnarray}
The vector and tensor indices are raised or lowered by $g^{\mu \nu}$ or $g_{\mu \nu}$,
and ${\cal P}_{\mu} e^{- i p_{\mu} \cdot x^{\mu}} = (p_{\mu} - qA_{\mu}) e^{- i p_{\mu} \cdot x^{\mu}}$.
For an electric field not perpendicular to a magnetic field, a Lorentz frame can be found in which both fields are parallel or
antiparallel and assumed to point along the $z$-direction. The space-dependent gauge field for a constant electromagnetic field is
\begin{eqnarray}
A_{\mu} = (- E z, 0, - Bx, 0).
\end{eqnarray}
The nonvanishing components of the field tensor $F_{\mu \nu} = \partial_{\mu} A_{\nu} - \partial_{\nu} A_{\mu}$ are
\begin{eqnarray}
F_{03} = -F_{30} = E, \quad F_{12} = - F_{21} = -B. \label{f ten}
\end{eqnarray}

The spin-1/2 fermions obey the Dirac equation
\begin{eqnarray}
(\gamma^{\mu} {\cal P}_{\mu} - m) \Psi = 0, \label{dir eq}
\end{eqnarray}
where $\{ \gamma^{\mu}, \gamma^{\nu} \} = 2 g^{\mu \nu}$ and in the standard representation
\begin{eqnarray}
\gamma^{0} =
\begin{pmatrix}I_2 & 0 \\
     0  & -I_2 \end{pmatrix}, \quad
      \gamma^{i} = \begin{pmatrix} 0 & \sigma_i \\
     - \sigma_i  & 0
     \end{pmatrix}.
\end{eqnarray}
The solution of Eq. (\ref{dir eq}) may be found from the solution for the squared Dirac operator
\begin{eqnarray}
(\gamma^{\mu} P_{\mu} - m) (\gamma^{\mu} P_{\mu} + m) = g^{\mu \nu} P_{\mu} P_{\nu} - m^2 - q \sigma^{\mu \nu} F_{\mu \nu}, \quad
\sigma^{\mu \nu}= \frac{i}{4} [\gamma^{\mu}, \gamma^{\nu}].
\end{eqnarray}
For the field tensor (\ref{f ten}), the squared Dirac equation becomes
\begin{eqnarray}
\Bigl[ g^{\mu \nu} P_{\mu} P_{\nu} - m^2 - 2 (q E) \sigma^{03} + 2(qB) \sigma^{12} \Bigr] \Phi = 0, \label{sq eq}
\end{eqnarray}
where
\begin{eqnarray}
\sigma^{03} = \frac{i}{2} \begin{pmatrix}
  0 & \sigma_z \\
     \sigma_z  & 0 \end{pmatrix}, \quad \sigma^{12} = \frac{1}{2} \begin{pmatrix}
   \sigma_z & 0\\
    0 & \sigma_z \end{pmatrix}. \label{spin ten}
\end{eqnarray}
The Klein-Gordon equation for scalar QED is the diagonal component of Eq. (\ref{sq eq}), neglecting terms involving $\sigma^{\mu \nu}$.
The following bispinors diagonalize both spin tensors (\ref{spin ten})
\begin{eqnarray}
\Xi_{(+)}^{\{+\}} = \frac{1}{\sqrt{2}} \begin{pmatrix}
1\\
0\\
1\\
0 \end{pmatrix}, \quad \Xi_{(+)}^{\{- \}} = \frac{1}{\sqrt{2}} \begin{pmatrix}
0\\
1\\
0\\
-1 \end{pmatrix}, \quad \Xi_{(-)}^{\{+\}} = \frac{1}{\sqrt{2}} \begin{pmatrix}
1\\
0\\
-1\\
0 \end{pmatrix}, \quad \Xi_{(-)}^{\{- \}} = \frac{1}{\sqrt{2}} \begin{pmatrix}
0\\
1\\
0\\
1 \end{pmatrix}. \label{bi sp}
\end{eqnarray}
The spin eigenvalues $(\pm1)$ of $\sigma^{03}$ will be denoted by $(r)$ whereas the spin eigenvalues
$\{\pm 1 \}$ of $\sigma^{12}$ by $2 \sigma$ along the magnetic field with $\sigma = \pm 1/2$.

\section{Jost Functions and Bogoliubov Coefficients} \label{sec jost}

The Klein-Gordon equation and the spin-diagonal Fourier component of the squared Dirac equation in electric or magnetic fields
pointing in one direction has the form of the Schr\"{o}dinger equation
\begin{eqnarray}
[\partial_{\mu}^2 \pm k^2 + V (x_{\mu}) ] \varphi_{k} (t) = 0.
\end{eqnarray}
Here for electric fields $\partial_{\mu} = \partial_t$, $+ k^2$, and $V \geq 0$  in the time-dependent gauge,
$\partial_{\mu} = \partial_z$, $ - k^2$, and $V \geq 0$ in the space-dependent gauge, and for magnetic fields
$\partial_{\mu} = \partial_x$, $+ k^2$, and $V \leq 0$. For localized electromagnetic fields $V(x_{\mu})$ approaches
to constant values at $x_{\mu} = \pm \infty$,  though $V(x_{\mu})$ may increase for electromagnetic fields that extend the whole space.
We shall introduce the Jost functions (for instance, see Refs. \cite{taylor,das}) and express the Bogoliubov coefficients in terms
of them. For the sake of simplicity, we assume $V ( \pm \infty) = {\rm constant}$ and consider the Klein-Gordon equation.

\subsection{Electric Field in Time-Dependent Gauge} \label{sec time-ga}

In the time-dependent gauge, the Fourier time-component Klein-Gordon equation takes the form
\begin{eqnarray}
[\partial_{t}^2 + \omega^2 + V (t) ] \varphi_{k} (t) = 0. \label{E-jos1}
\end{eqnarray}
As $V \geq 0$, Eq. (\ref{E-jos1}) is a scattering problem over the barrier. One may introduce the solutions with the asymptotic form
\begin{eqnarray}
f(t, \omega) \stackrel{t = \infty}{\longrightarrow} \frac{e^{-i \omega_{(+)}t}}{\sqrt{2 \omega_{(+)}}}, \quad
g(t, \omega) \stackrel{t = - \infty}{\longrightarrow} \frac{e^{i \omega_{(-)}t}}{\sqrt{2 \omega_{(-)}}}.
\end{eqnarray}
where $\omega^2_{(\pm)} = \omega^2 + V (\pm \infty)$ are the asymptotic frequencies in the future and in the past.
And $f(t,-\omega)$ and $g(t,-\omega)$ are another solutions independent of $f(t,\omega)$ and $g(t,\omega)$ in each asymptotic region.
Each set of solutions has the Wronskian
\begin{eqnarray}
[f(t,\omega), f(t, -\omega)] = i, \quad [g(t,\omega), g(t, -\omega)] = - i , \label{jos wr1}
\end{eqnarray}
where $[f, g] = f \partial_t g - g \partial_t f$. Hence $f(t,\omega)$ and $f(t, -\omega)$ form one set of independent solutions
while $g(t,\omega)$ and $g(t, -\omega)$ form another set.
So each set of solutions can be expressed in terms of the other set as
\begin{eqnarray}
f(t, \omega) = C_1 (\omega) g(t, -\omega) + C_2 (\omega) g(t, \omega), \nonumber\\
g(t, \omega) = \tilde{C}_1 (\omega) f(t, -\omega) + \tilde{C}_2 (\omega) f(t, \omega), \label{jos rel1}
\end{eqnarray}
where the Jost functions (coefficients) are determined as
\begin{eqnarray}
C_1 (\omega) &=& -i [f(t,\omega), g(t,\omega)] = \tilde{C}_1 (\omega), \nonumber\\
C_2 (\omega) &=& i [f(t,\omega), g(t,-\omega)] = - \tilde{C}_2 (-\omega). \label{bog coef1}
\end{eqnarray}
It holds that
\begin{eqnarray}
C_1 (-\omega) = C^{*}_1 (\omega), \quad C_2 (-\omega) = C^{*}_2 (\omega),
\end{eqnarray}
and
\begin{eqnarray}
|C_1 (\omega)|^2 - |C_2 (\omega)|^2 = 1.
\end{eqnarray}

From the Wronskian condition (\ref{jos wr1}), the set $\{ f(t,\omega) \}$ and another set  $\{f(t, -\omega) \}$ satisfy the orthonormality
\begin{eqnarray}
(f(t, \omega), f(t, \omega)) = 1, \quad (f(t, -\omega), f(t, -\omega)) = -1, \quad (f(t, \omega), f(t, -\omega)) = 0,
\end{eqnarray}
with respect to the inner product
\begin{eqnarray}
(\varphi_1, \varphi_2) = i [\varphi_1^*, \varphi_2] = i \int d^3 x \Bigl( \varphi_1^* \frac{\partial}{\partial t} \varphi_2 - \varphi_2 \frac{\partial}{\partial t} \varphi_1^* \Bigr).
\end{eqnarray}
Similarly, we have
\begin{eqnarray}
(g(t, \omega), g(t, \omega)) = -1, \quad (g(t, -\omega), g(t, -\omega)) = 1, \quad (g(t, \omega), g(t, -\omega)) = 0.
\end{eqnarray}
Denoting the positive frequency solutions $f(t, \omega)$ and $g(t, -\omega)$ by $\phi_{\rm (out)}^{(+)} (\omega)$ and $\phi_{\rm (in)}^{(+)} (\omega)$
and the negative frequency solutions $f(t, -\omega)$ and $g(t, \omega)$ by $\phi_{\rm (out)}^{(-)} (\omega)$ and $\phi_{\rm (in)}^{(-)} (\omega)$
and suppressing the momentum for simplicity, we may quantize the field in the future as
\begin{eqnarray}
\Phi (t,x) = \sum_{\omega} [ \phi_{\rm (out)}^{(+)} (\omega) a_{\rm out} (\omega) + \phi_{\rm (out)}^{(-)} (\omega) b_{\rm out}^{\dagger} (\omega)],
\end{eqnarray}
and in the past as
\begin{eqnarray}
\Phi (t,x) = \sum_{\omega} [ \phi_{\rm (in)}^{(+)} (\omega) a_{\rm in} (\omega) + \phi_{\rm (in)}^{(-)} (\omega) b_{\rm in}^{\dagger} (\omega)],
\end{eqnarray}
where $a_{\rm out}$ and $a_{\rm in}$ are the annihilation operators for particles and $b_{\rm out}^{\dagger}$ and $b_{\rm in}^{\dagger}$ are the
creation operators for antiparticles. Then the relations (\ref{jos rel1}) for Jost functions lead to
\begin{eqnarray}
\phi_{\rm (in)}^{(+)} (\omega) &=& C_1^* (\omega) \phi_{\rm (out)}^{(+)} (\omega) - C_2 (\omega) \phi_{\rm (out)}^{(-)} (\omega), \nonumber\\
 \phi_{\rm (in)}^{(-)} (\omega) &=& C_1 (\omega) \phi_{\rm (out)}^{(-)} (\omega) - C_2^* (\omega) \phi_{\rm (out)}^{(+)} (\omega). \label{time bog1}
\end{eqnarray}
Finally, we find the Bogoliubov transformation
\begin{eqnarray}
a_{\rm out} (\omega) &=& C_1^* (\omega) a_{\rm in} (\omega) - C_2^* (\omega) b^{\dagger}_{\rm in} (\omega), \nonumber\\
b_{\rm out}^{\dagger} (\omega) &=& C_1 (\omega) b_{\rm in}^{\dagger} (\omega) - C_2 (\omega) a_{\rm in} (\omega). \label{time bog2}
\end{eqnarray}
Hence the Jost functions (\ref{jos rel1}) determine the Bogoliubov coefficients
\begin{eqnarray}
\alpha_{\omega} = C^*_1 (\omega), \quad \beta_{\omega} = - C_2^* (\omega).
\end{eqnarray}

\subsection{Electric Field in Space-Dependent Gauge}\label{sec space-ga}

In the space-dependent gauge, the Fourier-component Klein-Gordon equation takes the form
\begin{eqnarray}
[\partial_{z}^2 - k^2 + V (z) ] \varphi_{k} (z) = 0, \label{E-jos2}
\end{eqnarray}
where $-k^2 + V \leq 0$ for some region of $z$ but $-k^2 + V(\pm \infty) = k^2_{(\pm)} \geq 0$. Thus Eq. (\ref{E-jos2}) becomes a tunneling problem
under the barrier. The asymptotic solutions now change as
\begin{eqnarray}
f(z, k) \stackrel{z = \infty}{\longrightarrow} \frac{e^{i k_{(+)}z}}{\sqrt{2 k_{(+)}}}, \quad
g(z, k) \stackrel{z = - \infty}{\longrightarrow} \frac{e^{-i k_{(-)}z}}{\sqrt{2 k_{(-)}}}. \label{jos asym2}
\end{eqnarray}
And $f(z,-k)$ and $g(z,-k)$ are independent solutions in each asymptotic region.
As each set of solutions has the Wronskian
\begin{eqnarray}
[f(z,k), f(z, -k)] = -i, \quad [g(z,k), g(z, -k)] = i, \label{jos wr2}
\end{eqnarray}
where $[f,g] = f \partial_z g - g \partial_z f$, each set of solutions can be expressed in terms of the other set
\begin{eqnarray}
f(z, k) = C_1 (k) g(z, -k) + C_2 (k) g(z, k), \nonumber\\
g(z, k) = \tilde{C}_1 (k) f(z, -k) + \tilde{C}_2 (k) f(z, k). \label{jos rel2}
\end{eqnarray}
The Jost functions are determined as
\begin{eqnarray}
C_1 (k) &=& i [f(z,k), g(z,k)] = \tilde{C}_1 (k), \nonumber\\
C_2 (k) &=& -i [f(z,k), g(z,-k)] = - \tilde{C}_2 (-k), \label{bog coef2}
\end{eqnarray}
and satisfy
\begin{eqnarray}
C_1 (-k) = C^{*}_1 (k), \quad C_2 (-k) = C^{*}_2 (k). \label{bog coef2-2}
\end{eqnarray}
Hence the relation holds:
\begin{eqnarray}
|C_1 (k)|^2 - |C_2 (k)|^2 = 1. \label{bog coef2-3}
\end{eqnarray}

In the space-dependent gauge, the field should be second quantized to resolve the Klein paradox from tunneling barrier.
In the St\"{u}ckelberg-Feynman picture \cite{feynman} for particle and antiparticle (see also Refs. \cite{nikishov70,damour,hansenravndal}),
$g(z, -k)$ multiplied with $e^{-i \omega t}$ describes a particle incoming from $z = - \infty$ while $f(z, k)$ multiplied
with $e^{-i (-\omega) (-t)}$ describes an antiparticle outgoing to $z = \infty$, traveling backward in time.
Since these wave functions correspond to the annihilation process of particle and antiparticle approaching from each asymptotic
region to an interaction region of electric field, we may associate the in-vacuum, $\phi_{\rm (in)}^{(+)} (\omega)$
and $\phi_{\rm (in)}^{(-)} (\omega)$, respectively. Similarly, $g(z, k)$ describes a particle outgoing
to $z = - \infty$ while $f(z, k)$ describes an antiparticle incoming from $z = \infty$, traveling backward in time.
This process is pair production and associates the out-vacuum, $\phi_{\rm (out)}^{(+)} (\omega)$
and $\phi_{\rm (out)}^{(-)}$, respectively. Then the field may be quantized as
\begin{eqnarray}
\Phi (t,x) = \sum_{k} [ g(z,k) a_{\rm out} (k) + f(z,-k) b_{\rm out}^{\dagger} (k)],
\end{eqnarray}
and
\begin{eqnarray}
\Phi (t,x) = \sum_{k} [ g(z,-k) a_{\rm in} (k) + f(z,k) b_{\rm in}^{\dagger} (k)].
\end{eqnarray}
Using the Jost functions (\ref{jos rel2}) and the relations (\ref{jos wr2}), (\ref{bog coef2}), (\ref{bog coef2-2}) and (\ref{bog coef2-3}),
we find the Bogoliubov transformations
\begin{eqnarray}
\phi^{(+)}_{\rm (in)} &=& C^*_1 (k) \phi^{(-)}_{\rm (in)} - C_2 (k) \phi^{(-)}_{\rm (out)}, \nonumber\\
\phi^{(-)}_{\rm (in)} &=& C_1 (k) \phi^{(+)}_{\rm (out)} + C_2 (k) \phi^{(+)}_{\rm (in)},
\end{eqnarray}
and
\begin{eqnarray}
a_{\rm out} &=& - \frac{C^*_1 (k)}{C^*_2 (k)} a_{\rm in} - \frac{1}{C^*_2 (k)} b^{\dagger}_{\rm in}, \nonumber\\
b_{\rm out} &=& \frac{C^*_1 (k)}{C_2 (k)} b_{\rm in} + \frac{1}{C_2 (k)} a^{\dagger}_{\rm in}. \label{space bog tran}
\end{eqnarray}
Thus the Bogoliubov coefficients are
\begin{eqnarray}
\alpha_k = - \frac{C^*_1 (k)}{C^*_2 (k)}, \quad \beta_k = - \frac{1}{C^*_2 (k)}. \label{space bog-fin}
\end{eqnarray}

\subsection{Magnetic Field in Space-Dependent Gauge} \label{sec mag-ga}

In magnetic fields the Klein-Gordon equation becomes a bounded system
\begin{eqnarray}
[\partial_{z}^2 - {\kappa}^2 + V (z) ] \varphi_{\kappa} (z) = 0, \label{B-jos3}
\end{eqnarray}
where $V \geq 0$ and $- {\kappa}^2 + V(\pm \infty) = - {\kappa}^2_{(\pm)}$. Instead of imposing bound states over the whole space,
we consider the asymptotically bounded behavior at $z = \infty$
\begin{eqnarray}
\varphi_{\kappa} (t) = T(\kappa) \frac{e^{- \kappa_{(+)} z}}{\sqrt{2 \kappa_{(+)}}}, \label{jos out3}
\end{eqnarray}
but a general form at $z = - \infty$
\begin{eqnarray}
\varphi_{\kappa} (z) = \frac{e^{\kappa_{(-)} z}}{\sqrt{2 \kappa_{(-)}}} + R(\kappa) \frac{e^{- \kappa_{(-)} z}}{\sqrt{2 \kappa_{(-)}}}. \label{jos in3}
\end{eqnarray}
In analogy with Secs. \ref{sec time-ga} and \ref{sec space-ga}, we may introduce the exponentially decreasing solutions in each asymptotic region
\begin{eqnarray}
f(z, \kappa) \stackrel{z = \infty}{\longrightarrow} \frac{e^{-\kappa_{(+)} z}}{\sqrt{2 \kappa_{(+)}}}, \quad
g(z, \kappa) \stackrel{z = - \infty}{\longrightarrow} \frac{e^{\kappa_{(-)} z}}{\sqrt{2 \kappa_{(-)}}}, \label{jos fn3}
\end{eqnarray}
and the exponentially increasing solutions $f(z,-\kappa)$ and $g(z,-\kappa)$ in their region as independent solutions. Each set of solutions
satisfies the Wronskian
\begin{eqnarray}
[f(z,\kappa), f(z, -\kappa)] = 1, \quad [g(z,\kappa), g(z, -\kappa)] = -1. \label{jos wr3}
\end{eqnarray}
We may then express each set of solutions in terms of the other set as
\begin{eqnarray}
f(z, \kappa) = C_1 (\kappa) g(z, \kappa) + C_2 (\kappa) g(z, - \kappa), \nonumber\\
g(z, \kappa) = \tilde{C}_1 (\kappa) f(z, \kappa) + \tilde{C}_2 (\kappa) f(z, -\kappa), \label{jos rel3}
\end{eqnarray}
where the Jost functions are determined from (\ref{jos wr3})
\begin{eqnarray}
C_1 (\kappa) &=& - [f(z,\kappa), g(z,-\kappa)] = - \tilde{C}_1 (-\kappa), \nonumber\\
C_2 (\kappa) &=&  [f(z,\kappa), g(z,\kappa)] = \tilde{C}_2 (-\kappa). \label{bog coef3}
\end{eqnarray}

It follows that
\begin{eqnarray}
C_2 (\kappa) C_2 (- \kappa) - C_1 (\kappa) C_1 (- \kappa) = 1, \label{jos bog3}
\end{eqnarray}
and
\begin{eqnarray}
T(\kappa) = \frac{1}{C_1 (\kappa)}, \quad R(\kappa) = \frac{C_2 (\kappa)}{C_1(\kappa)}. \label{jos amp3}
\end{eqnarray}
Then from Eqs. (\ref{jos rel3}) and (\ref{jos amp3}) we have the relation
\begin{eqnarray}
R(\kappa) R(-\kappa) - T(\kappa) T(-\kappa) = 1.
\end{eqnarray}
Finally, we may define the inverse scattering matrix as the ratio of the amplitude for exponentially increasing part to the amplitude
for the exponentially decreasing part, which is now given by
\begin{eqnarray}
{\cal M}_{\kappa} = R(\kappa) = \frac{C_2 (\kappa)}{C_1(\kappa)}. \label{in sc mat}
\end{eqnarray}
Comparing with the Bogoliubov coefficient (\ref{space bog-fin}) which is the scattering matrix
${\cal S}_{k} = - C_1^*(k)/C_2^*(k)$ for the tunneling problem in electric fields, Eq. (\ref{in sc mat})
is justified to be called the ``inverse scattering matrix'' for the bounded system.
The poles of ${\cal S}_{\kappa}$ for physically bound states correspond to the zeros for ${\cal M}_{\kappa}$
and the solution is exponentially decreasing in each asymptotic region.

\section{Formulas for Parabolic Cylinder Function, Gamma Function and Hypergeometric Function} \label{cyl fun}

To make the paper self-contained, we include the asymptotic formulas, the connection formulas for the parabolic cylinder function,
and the connection formula for the hypergeometric function. The asymptotic formulas 9.246 of Ref. \cite{gr-table} are
\begin{eqnarray}
D_p (z) &=& e^{- \frac{z^2}{4}} z^p, \quad (|{\rm arg }\, z| < \frac{3 \pi}{4}), \nonumber\\
D_p (z) &=& e^{- \frac{z^2}{4}} z^p - \frac{\sqrt{2 \pi}}{\Gamma(-p)} e^{i p \pi} e^{\frac{z^2}{4}} z^{-p -1}, \quad (\frac{\pi}{4} < {\rm arg } \, z < \frac{5 \pi}{4}), \nonumber\\
D_p (z) &=& e^{- \frac{z^2}{4}} z^p - \frac{\sqrt{2 \pi}}{\Gamma(-p)} e^{-i p \pi} e^{\frac{z^2}{4}} z^{-p -1}, \quad (- \frac{\pi}{4} > {\rm arg } \, z > \frac{5 \pi}{4}). \label{par asy}
\end{eqnarray}
The connection formulas 9.248 of Ref. \cite{gr-table} for the parabolic cylinder function are
\begin{eqnarray}
D_p (z) &=& \frac{\Gamma (p+1)}{\sqrt{2 \pi}} \Bigl[e^{ip \frac{\pi}{2}} D_{-p-1} (iz) + e^{-ip \frac{\pi}{2}} D_{-p-1} (-i z) \Bigr], \label{asym 1} \\
D_p (z) &=&  \sqrt{2 \pi} \frac{e^{-i(p+1) \frac{\pi}{2}}}{\Gamma (-p)} D_{-p-1} (i z) + e^{-ip \pi} D_{p} (- z), \label{asym 2} \\
D_{p} (z) &=& \sqrt{2 \pi} \frac{e^{i(p+1) \frac{\pi}{2}}}{\Gamma (-p)} D_{-p-1} (-i z) + e^{ip \pi} D_{p} (-z). \label{asym 3}
\end{eqnarray}

The formulas 8.332 of Ref. \cite{gr-table} for the gamma function are  
\begin{eqnarray}
|\Gamma (\frac{1}{2} + ix)|^2 = \frac{\pi}{\cosh (\pi x)}, \quad |\Gamma (ix)|^2 = \frac{\pi}{x \sinh (\pi x)}, \quad
|\Gamma (1 + ix)|^2 = \frac{\pi x}{\sinh (\pi x)}, \label{gamma form1}
\end{eqnarray} 
and the identity 8.334 of Ref. \cite{gr-table} is
\begin{eqnarray}
\Gamma (1-x) \Gamma (x) = \frac{\pi}{\sin(\pi x)}. \label{gamma form2}
\end{eqnarray}

The connection formula 9.132 of Ref. \cite{gr-table} for the hypergeometric function is
\begin{eqnarray}
F(\alpha, \beta; \gamma; z) &=& \frac{\Gamma (\gamma) \Gamma(\beta - \alpha)}{\Gamma (\beta) \Gamma(\gamma - \alpha)} (-z)^{- \alpha} F(\alpha, \alpha+1 - \gamma; \alpha + 1 - \beta; \frac{1}{z}) \nonumber\\&& + \frac{\Gamma (\gamma) \Gamma(\alpha - \beta)}{\Gamma (\alpha) \Gamma(\gamma - \beta)} (-z)^{- \beta} F(\beta, \beta+1 - \gamma; \beta + 1 - \alpha; \frac{1}{z}). \label{hyper con}
\end{eqnarray}


\begin{thebibliography}{99}

\bibitem{heisenbergeuler} W.~Heisenberg and H.~Euler, Z. Phys. {\bf 98}, 714 (1936); an English translation is available at physics/0605038.

\bibitem{weisskopf} V.~Weisskopf, K. Dan. Vidensk. Selsk. Mat. Fy. Medd. {\bf 14}, 1 (1936); reprinted in {\it
Quantum Electrodynamics}, edited by J. Schwinger (Dover, New York, 1958).

\bibitem{schwinger} J.~Schwinger, Phys. Rev. {\bf 82}, 664 (1951).

\bibitem{adler} S.~L.~Adler, Ann. Phys. {\bf 67}, 599 (1971).

\bibitem{erber} T.~Erber, Rev. Mod. Phys. {\bf 38}, 626 (1966).

\bibitem{klni} J.~J.~Klein and B.~P.~Nigam, Phys. Rev. {\bf 135}, 1279 (1964).

\bibitem{dittrichreuter} W.~Dittrich and M.~Reuter, {\it Effective Lagrangians in Quantum Electrodynamics},
Lect. Notes Phys. {\bf 220} 1 (Springer, Berlin, 1985).

\bibitem{fgs} E.~S.~Fradkin, D.~M.~Gitman, and S.~M.~Shvartsman, {\it  Quantum Electrodynamics with Unstable Vacuum},
(Springer, Berlin, 1991).

\bibitem{dittrichgies} W.~Dittrich and H.~Gies, {\it Probing the Quantum Vacuum}, Springer Tracts Mod. Phys. {\bf 166}, 1, (Springer, Berlin, 2000).

\bibitem{schubert-rev} C.~Schubert, Phys. Rept. {\bf 355}, 73 (2001).

\bibitem{dunne-rev} G.~V.~Dunne, ``Heisenberg-Euler Effective Lagrangians: Basics and Extensions,'' in {\it From Fields to Strings:
Circumnavigating Theoretical Physics}, Vol. I, pp 445, M.A. Shifman et al. (Eds.), (World Scientific, Singapore, 2004) [hep-th/0406216].

\bibitem{ringwald} A. Ringwald, ``Fundamental physics at an x-ray free electron laser,''
in {\it Proceedings of Erice Workshop On Electromagnetic Probes Of Fundamental Physics}, edited by W.~Marciano and S.~White,
(World Scientific, Singapore, 2003) [hep-ph/0112254].

\bibitem{mash} M.~Marklund and P.~K.~Shukla, Rev. Mod. Phys. {\bf 78}, 591 (2006).

\bibitem{dunne-work} G.~V.~Dunne, Eur. Phys. J. D {\bf 55}, 327 (2009).

\bibitem{behpt} S.~V.~Bulanov, T.~Zh.~Esirkepov, D.~Habs, F.~Pegoraro, and T.~Tajima, Eur. Phys. J. D {\bf 55}, 483 (2009).

\bibitem{ruffini} R.~Ruffini, G.~Vereshchagin, and S-S.~Xue, Phys. Rept. {\bf 487}, 1 (2010).

\bibitem{blviwi} S.~Blau, M.~Visser, and A.~Wipf, Int. J. Mod. Phys.  {\bf A 6}, 5409 (1991) .

\bibitem{eorbz} E.~Elizalde, S.~D.~Odintsov, A.~Romeo, A.~A.~Bytsenk, and S.~Zerbini, {\it Zeta Regularization Techniques with Applications},
(World Scientific, Singapore, 1994).

\bibitem{rescsc} M.~Reuter, M.~G.~Schmidt, and C.~Schubert, Ann. Phys. (N.Y.) {\bf 259}, 313 (1997).

\bibitem{chd} D.~Cangemi, E.~D'Hoker, and G.~V.~Dunne, Phys. Rev. D {\bf 52}, 3163 (1995).

\bibitem{dunne-hall98} G.~V.~Dunne and T.~M.~Hall, Phys. Rev. D {\bf 58}, 105022 (1998).

\bibitem{dunne-hall98-2} G.~V.~Dunne and T.~M.~Hall, Phys. Lett. B {\bf 419}, 322 (1998).

\bibitem{fried} H.~M.~Fried and R.~P.~Woodard, Phys. Lett. B {\bf 524}, 233 (2002).

\bibitem{stuckelberg} E.~C.~G.~St\"{u}ckelberg, Helv. Phys. Acta {\bf 14}, 588 (1941).

\bibitem{feynman} R.~P.~Feynman, Phys. Rev. {\bf 76}, 749 (1949); {\it ibid} {\bf 74}, 936 (1948).

\bibitem{nikishov} A.~I.~Nikishov, Zh. Eksp. Teor. Fiz. {\bf 57}, 1210 (1969)  [Sov. Phys. JETP {\bf 30},  660 (1970)].

\bibitem{nikishov70} A.~I.~Nikishov, Nucl. Phys. B {\bf 21}, 346 (1970).

\bibitem{nikishov79} A.~I.~Nikishov, Tr. Fiz. Inst. im P. N. Lebedeva, Akad. Nauk SSSR {\bf 111}, 152 (1979)
[J. Sov. Laser Res. {\bf 6}, 619  (1985)].

\bibitem{damour} T.~Damour, ``Klein Paradox and Vacuum Polarization,`` in {\it Proceedings of the First Marcel Grossmann Meeting on General Relativity}, edited by R.~Ruffini
(North-Holland, Amsterdam, 1977) pp. 459.

\bibitem{hansenravndal} A.~Hansen and F.~Ravndal, Phys. Scr. {\bf 23}, 1036 (1981).

\bibitem{bmps} R.~Brout, S.~Massar, R.~Parentani, and Ph.~Spindel, Phys. Rept. {\bf 260}, 329 (1995).

\bibitem{gagigo} S.P. Gavrilov, D.M. Gitman, and A.E. Goncalves, J. Math. Phys. {\bf 39}, 3547 (1998).

\bibitem{gavgit} S.~P.~Gavrilov and  D.~M.~Gitman, Phys. Rev. {\bf D 53}, 7162 (1996).

\bibitem{gasp} C.~Gabriel and P.~Spindel, Ann. Phys. {\bf 284}, 263 (2000).

\bibitem{ggt} S.~P.~Gavrilov, D.~M.~Gitman, and J.~L.~Tomazelli, Nucl. Phys. B {\bf 795}, 645 (2008).

\bibitem{tanji} N.~Tanji, Ann. Phys. {\bf 324}, 1691 (2009).

\bibitem{fgl} K.~Fukushima, F.~Gelis, and T.~Lappi, Nucl. Phys. A {\bf 831}, 184 (2009).

\bibitem{soldati} R.~Soldati, J. Phys. A {\bf 44}, 305401 (2011).

\bibitem{schwinger51} J.~Schwinger, Proc. Natl Acad. Sci. U.S.A. {\bf 37}, 452 (1951) .

\bibitem{dewitt} B.~S.~DeWitt, Phys. Rept. {\bf 19}, 295 (1975) .

\bibitem{dewitt-book} B.~S.~DeWitt {\it The Global Approach to Quantum Field Theory Vol. 1, Vol. 2}, (Oxford University Press, New York, 2003).

\bibitem{ahn} J.~Ambj{o}rn, R.~J.~Hughes, and N.~K.~Nielsen, Ann. Phys. {\bf 150}, 92 (1983).

\bibitem{nikishov03} A.~I.~Nikishov, Zh. Eksp. Teor. Fiz. {\bf 123}, 211 (2003)  [Sov. Phys. JETP {\bf 96}, 180 (2003)].

\bibitem{kly08} S.~P.~Kim, H.~K.~Lee, and Y.~Yoon, Phys. Rev. D {\bf 78}, 105013 (2008).

\bibitem{kim09} S.~P.~Kim, AIP. Conf. Proc. {\bf 1150}, 95 (2009).

\bibitem{kly10} S.~P.~Kim, H.~K.~Lee, and Y.~Yoon, Phys. Rev. D {\bf 82}, 025015 (2010).

\bibitem{kly10-2} S.~P.~Kim, H.~K.~Lee, and Y.~Yoon, Phys. Rev. D {\bf 82}, 025016 (2010).

\bibitem{taylor} J.~R.~Taylor, {\it Scattering Theory}, (Dover Publications, New York, 2000).

\bibitem{chopak} Y.~M.~Cho and D.~G.~Pak, Phys. Rev. Lett. {\bf 86}, 1947 (2001).

\bibitem{bcp} W.~S.~Bae, Y.~M.~Cho, and D.G. Pak, Phys. Rev. D {\bf 64}, 017303 (2001).

\bibitem{gr-table} I.~S.~Gradshteyn and I.~M.~Ryzhik, {\it Table of Integrals, Series,
and Products}, (Academic Press, San Diego, 1994).

\bibitem{riemann} In the sheet $- \pi \leq {\rm arg z} < \pi$, 
$i = e^{i \pi/2}$, $1 = e^{i 0 \pi}$, $- i = e^{- i \pi/2}$, and $-1 = e^{-i \pi}$, so $\sqrt{-1} = e^{- i \pi/2} = -i$.

\bibitem{popovmarinov} V.~S.~Popov and N.~S.~Marinov, J. Nucl. Phys. (USSR) {\bf 16}, 809 (1972).

\bibitem{kim10} S.~P.~Kim, ``Vacuum Structure of de Sitter Space,'' [arXiv:1008.0577[hep-th]].

\bibitem{kim11} S.~P.~Kim, ``Probing the Vacuum Structure of Spacetime,'' [arXiv:1102.4154[hep-th]].

\bibitem{bunkintugov} F.~V.~Bunkin and I.~I.~Tugov, Sov. Phys. Dokl. {\bf 14}, 678 (1970).

\bibitem{daughertylerche} J.~K.~Daugherty and I.~Lerche, Phys. Rev. D {\bf 14}, 340 (1976).

\bibitem{popov} V.~S.~Popov, Sov. Phys. JETP {\bf 34}, 709 (1972); {\it ibid} {\bf 35}, 659 (1972).

\bibitem{kimpage02} S.~P.~Kim and D.~N.~Page, Phys. Rev. D {\bf 65}, 105002 (2002).

\bibitem{kimpage06} S.~P.~Kim and D.~N.~Page, Phys. Rev. D {\bf 73}, 065020 (2006).

\bibitem{greiner-book} W.~Greiner, {\it Relativistic Quantum Mechanics}, (Springer, Berlin, 1997).

\bibitem{das} A.~Das, {\it Integrable Models}, (World Scientific, Singapore, 1989).

\end{thebibliography}
\end{document}